**Conference paper:**
A Study of Three Dimensional Bubble Velocities at Co-current Gas-liquid Vertical Upward Bubbly Flows

**Authors:**
Hadiyan Yusuf Kuntoro; Manuel Banowski; Deendarlianto

**Publication date:**
4 December 2014

**Conference:**
The 1st International Conference on Engineering Technology and Industrial Application (ICETIA 2014)

**Venue:**
Universitas Muhammadiyah Surakarta, Surakarta, Indonesia

**Conference date:**
4 December 2014

**Pages:**
417-425

# A Study of Three Dimensional Bubble Velocities at Co-current Gas-liquid Vertical Upward Bubbly Flows

Hadiyan Yusuf Kuntoro[1, 2] *, Manuel Banowski[2], Deendarlianto[1]

[1]*Department of Mechanical and Industrial Engineering, Faculty of Engineering, Gadjah Mada University, Jalan Grafika 2 Yogyakarta 55281, Indonesia.*

[2]*Institute of Fluid Dynamics, Helmholtz-Zentrum Dresden-Rossendorf, Bautzner Landstraße 400, 01328 Dresden, Germany.*

*Corresponding author: hadiyan.y.kuntoro@mail.ugm.ac.id; h.kuntoro@hzdr.de

*Abstract* — Recently, experimental series of co-current gas-liquid upward bubbly flows in a 6 m-height and 54.8 mm i.d. vertical titanium pipe had been conducted at the TOPFLOW thermal hydraulic test facility, Helmholtz-Zentrum Dresden-Rossendorf, Germany. The experiments were initially performed to develop a high quality database of two-phase flows as well as to validate new CFD models. An ultrafast dual-layer electron beam X-ray tomography, named ROFEX, was used as measurement system with high spatial and temporal resolutions. The gathered cross sectional grey value image results from the tomography scanning were reconstructed, segmented and evaluated to acquire gas bubble parameters for instance bubble position, size and hold-up. To assign the correct paired bubbles from both measurement layers, a bubble pair algorithm was implemented on the basis of the highest probability values of bubbles in position, volume and velocity. Hereinafter, the individual characteristics of bubbles were calculated include instantaneous three dimensional bubble velocities, Sauter mean diameters and the movement angle in polar-azimuthal directions. The instantaneous three dimensional bubble velocities are discussed through the distribution results of the axial, horizontal, radial and azimuthal velocities in statistical parameters. The present results show satisfactory agreement with previous works.

## I. Introduction

Two-phase bubble flow is encountered in a wide range of industrial plants such as bubble column reactors, gas-liquid pipeline systems, and steam generators. This flow regime belongs to important phenomena which have to be understood in order to improve the effectiveness of the heat and mass transfer process as well as to perform safety analysis relating to the human and environment.

As the performance of modern computer increase throughout the decades, CFD simulations are extensively employed in industrial application to compute complex two-phase flow problems. The simulation results can provide valuable information for the optimization of industrial processes. Therefore, the validations of two-phase CFD models using experimental data with high spatial and temporal resolution are inevitable.

Numerous researchers developed some methods and techniques to investigate two-phase bubbly flow phenomena. Some examples are wire-mesh sensor [1-3], needle probes [4-6], electrical capacitance tomography [7-9] and X-ray CT (computed tomography) [10-12]. Wire-mesh sensor and needle probes are intrusive measurement techniques. Electrical capacitance and X-ray computed tomography are belong to the non-intrusive methods [13] but these techniques still do not reach the required temporal and spatial resolution at the same time [14].

At the Institute of Fluid Dynamics, Helmholtz-Zentrum Dresden-Rossendorf (HZDR), Germany, a novel measurement system was developed to observe gas-liquid phenomena in high spatial and temporal resolutions, namely **Ro**ssendorf **F**ast **E**lectron Beam **X**-ray Tomography (ROFEX). Instead using mechanical rotation of the scanner components as in conventional X-ray CT scanners, an electron beam is swept rapidly across a target using deflection coils. Hence, a frame rate of up to 8000 frames per second can be attained. For the velocity measurement purpose, the target and detector rings are dual constructed with a vertical distance of 10.2 mm.

In the case of bubbly flow, the implementation of an accurate method to evaluate the bubble velocity plays an important role. Commonly, for dual-layer measurement system, i.e. wire-mesh sensor and ROFEX, a cross-correlation method is applied [1, 2]. However, this method assumes that all bubbles present in a given time interval have the same axial velocities, depending on the radial position [15]. Hence, the individual instantaneous bubble velocity cannot be derived. Another method that can be taken is by detecting the time delay of each bubble passes through two measurement layers with known vertical distance. Then, the individual instantaneous bubble rise velocity can be calculated. To do this, an algorithm to assign the correct paired bubbles from both measurement layers must be implemented.

Patmonoaji et al. [16] had developed an algorithm to assign the correct paired bubbles from dual-layer ROFEX measurement database, called bubble pair algorithm. The present work deals with bubble velocity calculations in axial, horizontal, radial, and azimuthal directions from the bubble pair algorithm's results, as well as the bubble movement angle in polar and azimuthal directions. To check the reliability of a bubble pair algorithm, the validation of the velocity calculation results must be performed. The present goal is to meet a good agreement between velocity calculations from the bubble pair algorithm results with the established data and theory.

## II. EXPERIMENTAL SETUP

### A. The Ultrafast X-ray Tomography

The ultrafast dual-layer electron beam X-ray tomography ROFEX at the HZDR is a non-intrusive measurement system with high spatial and temporal resolutions. Fig. 1 shows the operating principle of the ultrafast dual-layer electron beam X-ray tomography ROFEX. In an operation of the ROFEX scanner, an electron beam of sufficient energy is produced by an electron beam gun, which is focused onto a 240° circular target. Fischer et al. [17] implied that the angle of 240° provides sufficient data for a complete tomography reconstruction. The electron beam is swept across the circular target by means of an electromagnetic deflection coils. The circular target deflects the electron beam, yielding an X-ray fan in the form of radiation that passes the investigation objects. The investigation objects attenuate the incoming radiation regarding the Lambert's Law. Here, the attenuation signals in form the radiation intensity signals are recorded by a fast X-ray detector system. The maximum scan frequency of ROFEX is 8000 frames per second. For the determination of bubbles velocities, ROFEX is constructed with two target and detector rings, which are able to scan the pipe cross section in two measuring layer with a vertical distance of 10.2 mm. Detail explanation about ROFEX had presented by Fischer et al. [17] and Fischer and Hampel [18].

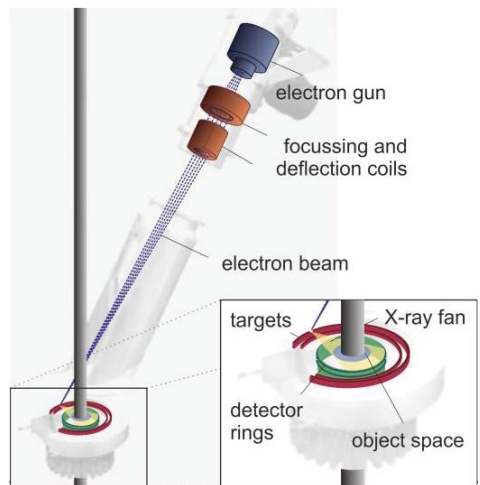

Fig. 1. The operating principle of ROFEX.

### B. TOPFLOW Test Facility

The gas-liquid flow experimental series were carried out in the TOPFLOW test facility at Institute of Fluid Dynamics in the Helmholtz-Zentrum Dresden-Rossendorf, Germany. A vertical titanium pipe with inner diameter of 54.8 mm and total length of about 6 m was used. In difference to a steel pipe, the utilization of titanium pipe is intended to reduce the wall thickness down to 1.6 mm for high pressure and high temperatures experimental conditions, but with less X-ray attenuation.

In the present experiments, air and water were used as fluids for co-current gas-liquid vertical upward bubbly flow. The superficial gas and liquid velocities were varied between 0.0025 m/s to 0.0898 m/s and 0.0405 m/s to 1.611 m/s, respectively. The pressure at top of the pipe was adjusted at 0.4 MPa, whilst the two-phase mixture was maintained at 30℃ ± 1 K. The gas injection device contains of annuli of cannulas which are fed with gas from two ring chambers. The cannulas have an inner diameter of 0.8 mm. The distance between ROFEX measurement layer and gas injection was varied from L/D = 0.1 to L/D = 60. Therefore, the evolution of gas-liquid flows is observable. Each flow regime in each specific L/D position is scanned for 10 s. The schematic view of the experimental apparatus is shown in Fig. 2.

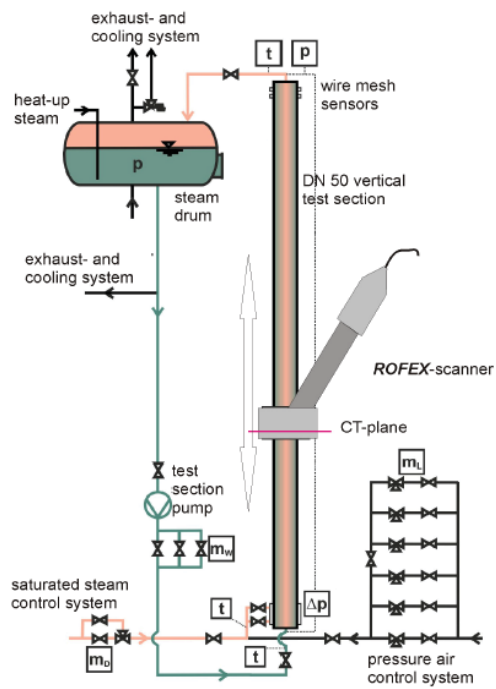

Fig. 2. The schematic view of experimental apparatus.

## III. DATA PROCESSING

### A. Image Reconstruction

Once a scanning process has been performed and data has been transferred to the PC, data is pre-processed and reconstructed using the filtered back projection, mentioned e.g. in Kak and Slaney [19]. The reconstructed data yield a stack of cross-sectional grey value images that represent the distribution of the gas and liquid. The cross sectional size of a frame is 108x108 pixel with spatial resolution 0.5x0.5 mm for each square pixel. The array length depends on the measuring time and chosen frequency, for example, for measuring time 10 s and frequency 2.5 kHz, the array length is 25 000 frames. Fig. 3 gives an example of an array data.

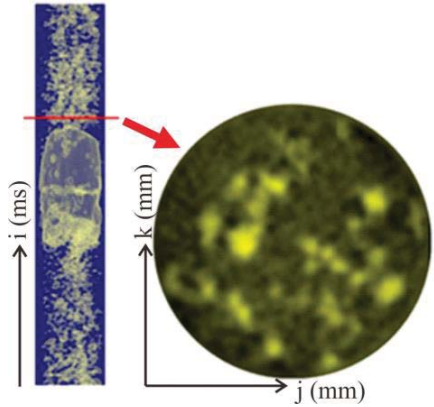

Fig. 3. A reconstructed tomography data (left: An array data; right: A cross-sectional frame).

*B. Images Segmentation*

To obtain the bubble parameters, the grey value arrays have to be binarized. The new segmentation algorithm for the grey value arrays had been developed by Banowski et al. [20] on the basis of the bubbles detection by pixel agglomeration. The idea is growing of bubble regions by beginning at the maximum grey values. The process is starting from maximum grey value until the grey value, which represents the signal-noise-ratio, equal 1 in the water phase. After pixel agglomeration in shrinking steps, the segmented bubbles regions are larger than the real bubbles. So, to obtain real bubbles sizes, an individual bubble threshold depending on their maximum grey value is used to cut wrong assigned pixels to the water phase. Further explanation about the image segmentation algorithm had been presented by Banowski et al. [20]. The example of segmented image is shown in Fig. 4, whereas a flowchart of the image segmentation algorithm is shown in Fig. 5.

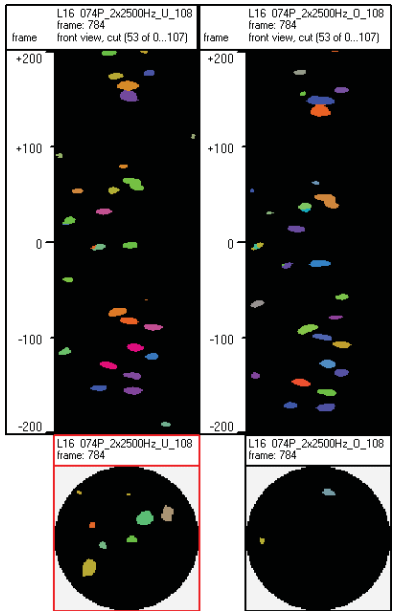

Fig. 4. The segmented image.

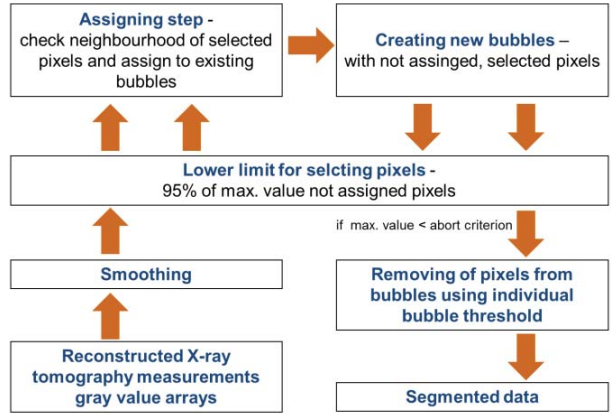

Fig. 5. Flowchart of the new image segmentation algorithm [20].

After the binarization is carried out, bubble parameters such as bubble virtual size, bubble position and bubble detected time are able to be derived. Bubble virtual size is the size dimension of a bubble from ROFEX measurement with the unit volume $mm^2 ms$. This is because the *i*-coordinate of an array data is in time coordinate (see Fig. 3).

*C. Bubble Pair Algorithm Process*

To evaluate the individual instantaneous bubble velocities from ROFEX measurement database, instead using cross-correlation methods, a bubble pair algorithm is implemented. This algorithm assigns the bubbles from both ROFEX measurement layers to related bubble pairs. The determination of correct paired bubbles is based on the highest total probability values of bubbles in position, volume and expected velocity. During the process, when the algorithm has found the most probable pair, both bubbles at different measurement layers are labelled as solved bubble to prevent multiple pair solution. The definite probability criteria for compared parameters in a bubble pair algorithm had been created by Patmonoaji et al. [16].

*D. Bubble Velocity Calculation*

After performing the bubble pair algorithm, the evaluation of three-dimensional bubble velocities can be carried out. Fig. 6 shows the velocity components that can be calculated after the correct paired bubbles are identified. The bubbles position data from ROFEX measurement database are stated in Cartesian coordinate, with the origin that is showed in Fig. 6. In the present work, the bubbles' centre of mass is used as the reference for velocity calculation. The calculation formula for each component is explained as follow:

*1) Axial Velocity*: this velocity component shows the bubbles movement rate in axial direction, which is calculated using (1).

$$U_{axial} = \Gamma/(i_O - i_U) \qquad (1)$$

$\Gamma$ is the distance between two measurement layers. For ROFEX measurement layers, the value is 10.2 mm. $i_O$ and $i_U$ is the time detection of paired bubbles in upper and lower measurement layers, respectively.

*2) Radial Velocity*: this velocity component expresses the bubbles movement rate in the radial direction. If the bubbles move toward the pipe wall, the radial velocity is negative, otherwise positive. This can be calculated using (2).

$$U_{radial} = \Delta rad/(i_O - i_U) \quad (2)$$

where
$$\Delta rad = -(r_O - r_U)$$
$$r_O = \sqrt{(j_O - R)^2 + (k_O - R)^2}$$
$$r_U = \sqrt{(j_U - R)^2 + (k_U - R)^2}$$

The values of $j$ and $k$ are the Cartesian coordinate positions of the bubbles in cross-sectional image, where subscripts $o$ and $u$ are symbols for bubbles in upper and lower layers, respectively. $R$ is the radius of the inner diameter pipe.

*3) Azimuthal Velocity*: The bubble movement rates in azimuthal direction are stated by this velocity component. The bubbles will have negative azimuthal velocity when the bubbles move in clockwise direction, otherwise positive, with the reference an observer is looking towards the oncoming co-current upward flow. This is stated in (3).

$$U_{azimuthal} = r_U \left[ \beta/(i_O - i_U) \right] \quad (3)$$

$$\beta = \cos^{-1}\left(\frac{j_O \cdot j_u + k_O \cdot k_u}{\sqrt{j_O^2 + k_O^2} \cdot \sqrt{j_U^2 + k_U^2}}\right) \quad (4)$$

The value $r_U$ is the radial distance of the detected bubbles from the centre of pipe in lower layer, whilst $\beta$ is the azimuthal angle difference between the paired bubbles (see Fig. 6) and calculated using (4).

*4) Horizontal Velocity*: this velocity component is a superposition between radial and azimuthal velocities. This is expressed in (5). The calculation of this velocity is based on the difference point in horizontal position according to the Cartesian coordinate.

$$U_{horizontal} = \left(\sqrt{(j_O - j_U)^2 + (k_O - k_U)^2}\right)/(i_O - i_U) \quad (5)$$

*5) 3D Velocity*: this velocity component is the resultant velocity between axial and horizontal velocities. This can be calculated using (6).

$$U_{3D} = \sqrt{U_{axial}^2 + U_{horizontal}^2} \quad (6)$$

By knowing the axial and horizontal velocities, the polar angle is also can be calculated by using (7).

$$\alpha = \tan^{-1}(U_{horizontal}/U_{axial}) \quad (7)$$

The size of the bubbles is generally stated in the Sauter mean diameter. To obtain the Sauter mean diameter of the bubbles, the actual volume of each bubble must be known. To do this, each bubble virtual size is multiplied by the corresponding axial velocity of the bubbles, then the Sauter mean diameter can be calculated using (8). $V_b$ is the bubble actual volume.

$$d_{SM} = \sqrt[3]{6V_b/\pi} \quad (8)$$

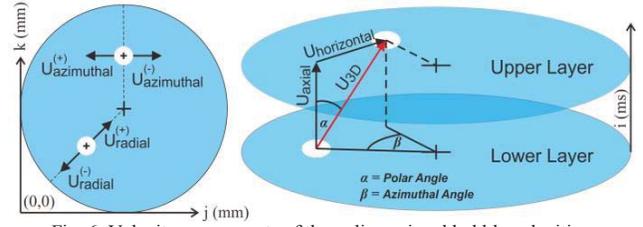

Fig. 6. Velocity components of three dimensional bubble velocities.

## IV. RESULTS AND DISCUSSIONS

### A. Results Verification

The reliability of a bubble pair algorithm from Patmonoaji et al. [16] can be evaluated by comparing the bubble velocity calculation's results to other established experimental database and theory. Experimental database from Hibiki et al. [4], Doup et al. [5], Tian et al. [21], and Lucas and Mishra [6] were used as the comparisons, whereas the power law velocity profile theory from Zuber and Findlay [22] was used. It should be noted that the experimental condition from the aforesaid authors are not exactly the same, but similar with the present work, i.e. the superficial gas-liquid velocities, inner-diameter pipe, and L/D.

Hibiki et al. [4], Doup et al. [5], Tian et al. [21] and Lucas and Mishra [6] employed the probe measurement technique to acquire the bubble parameters. Doup et al. [5] performed the experiments of co-current air-water vertical upward bubbly flow in a 50 mm inner-diameter pipe, while Hibiki et al. [4] and Tian et al. [21] used a 50.8 mm inner-diameter pipe. The larger inner-diameter pipe of 80 mm was used by Lucas and Mishra [6], which was originally used to develop the measurement technique for axial, azimuthal, and radial bubble velocities by using probe in the swirl and non-swirl condition.

Fig. 7 shows the development of the bubble axial velocity profiles from L/D = 8 to 60, with various $J_L$ at a constant $J_G$ = 0.0096 m/s. The present results are consistent with Hibiki et al's [4] experimental database which state that the introduction of bubbles into the liquid flow at $J_L \leq 1$ m/s flattened the velocity profile, with a relatively steep decrease close to the wall. For $J_L > 1$ m/s, the velocity profile came to be the power law distribution profile as the flow developed.

Fig. 8 shows the comparison of the time-averaged axial velocity in radial distribution between Doup et al. [5] (left box), Zuber and Findlay [22] theory and the present data (right box). In Doup et al's [5] data, the superficial gas velocities ($J_G$) are 0.09 m/s at L/D = 10 and 0.10 m/s at L/D = 32, with the constant superficial liquid velocity ($J_L$) of 1.08 m/s. The present test matrix is 1.017 m/s for $J_L$ and 0.0898 m/s for $J_G$, at L/D = 60, 23 and 8. The present data show a good agreement velocity profile with Zuber and Findlay [22] theory. For the comparison with Doup et al's [5] data, the present velocity profile shows a similar trend. The little disagreement of velocity profile occurs in the radial distribution near the wall (r/R = 0.5 - 1). It is supposed that the little bit higher $J_L$ for Doup et al's [5] data become the reason of the higher estimation of velocity profile in this range.

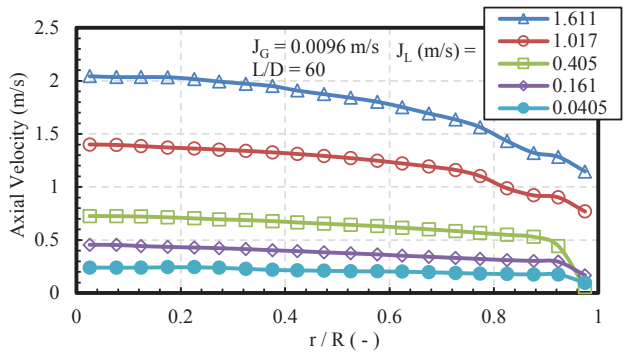
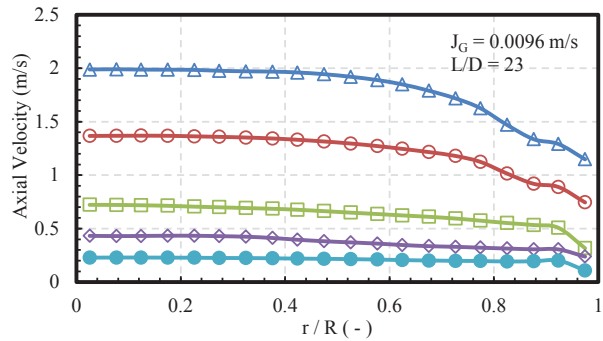
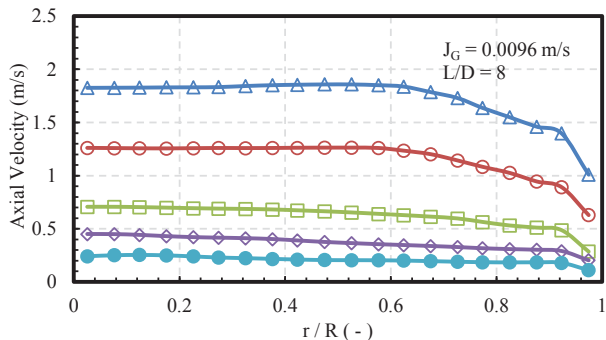

Fig. 7. The development of the bubble axial velocity profiles.

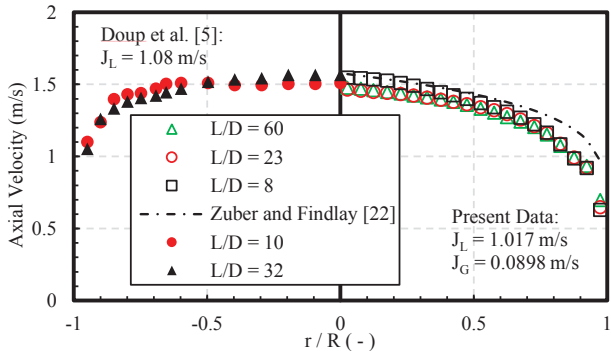

Fig. 8. The comparison of the time-averaged axial velocity in radial distribution.

Fig. 9 shows the comparison of the time-averaged Sauter mean diameter in radial distribution between Tian et al's [21] data and the present data. The superficial gas velocities between them is similar, $J_G = 0.009$ m/s for Tian et al. [21] and $J_G = 0.0096$ m/s for present data. The superficial liquid velocities of Tian et al. [21] is 0.071 m/s with L/D = 22, and for present data is 0.0405 m/s with L/D = 23. From r/R = 0 – 0.8, a good agreement is attained, while for r/R = 0.8 – 1, a slight difference is noted. Start from r/R = 0.8 to the pipe wall, an increasing trend of Sauter mean diameter was observed by Tian et al. [21], whilst the present data show a decreasing trend. It is expected that the probe decrease the bubble velocity near the wall due to its intrusive measurement characteristic. Then, this results in longer measurement of the bubble chord length.

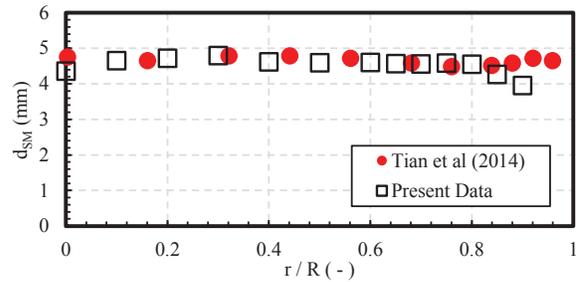

Fig. 9. The comparison of the time-averaged Sauter mean diameter in radial distribution.

The present calculated time-averaged radial and azimuthal velocities are compared to Lucas and Mishra's [6] experimental data, as shown in Fig. 10 and 11. For Fig. 10 and 11, the present flow regime has $J_L = 0.405$ m/s and $J_G = 0.0096$ m/s at L/D = 23. Lucas and Mishra's [6] flow regime has $J_L = 0.41$ m/s at L/D = 25 with $J_G = 0.068$ m/s for Fig. 10, and $J_G = 0.037$ for Fig. 11. A high discrepancy of the time-averaged radial and azimuthal velocities are exhibited between Lucas and Mishra's [6] data and the present data. The relatively higher inner diameter pipe than the present work is presumed to be the cause. A higher inner diameter pipe enables the bubbles to move relatively free toward any direction due to the smaller bubbles density rather than in the smaller inner diameter pipe, resulting the higher radial and azimuthal movement. Moreover, due to the presence of a probe in the flow field, bubbles approaching the probe tip are disturbed, even deformed, and slowed down, causing a change in the bubble orientation direction during the bubble-sensor touching process [23]. Therefore, the bubbles movement directions in the Lucas and Mishra's [6] experiments are not purely due to the interplay of the lateral force between gas, liquid and pipe wall, but also including the interaction with the probe. This proves that the present experiments show the better results because the implementation of non-intrusive measurement systems.

The negative and positive values of the radial and azimuthal velocities represent the direction of the bubbles movements. The negative value on the radial velocity implies that the bubbles move toward the pipe wall, and vice versa. For the azimuthal velocity, the negative value implies the bubbles movements in the clockwise direction (when the flow is moving co-currently upwards along the pipe towards the observer), and vice versa.

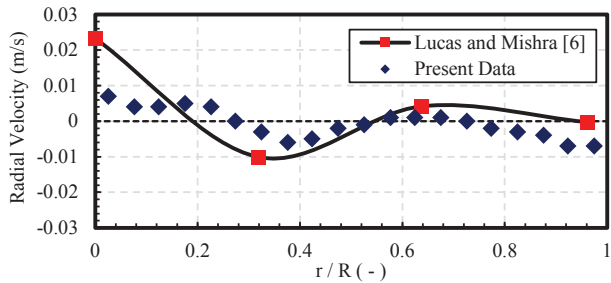

Fig. 10. The comparison of the time-averaged radial velocity in radial distribution.

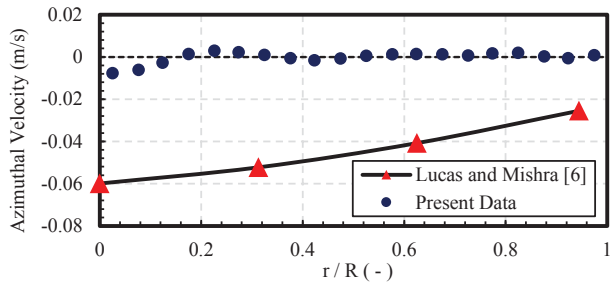

Fig. 11. The comparison of the time-averaged azimuthal velocity in radial distribution.

*B. Interpretation for the Lateral Forces*

In vertical pipe flow, the development pattern of the local radial and azimuthal velocities are affected by the interplay of the bubble lateral forces, which act perpendicularly to the flow direction (non-drag forces). The forces taken into account are lift force, wall force and dispersion force. A number of correlations for these bubble forces can be found in literature [24-26].

Fig. 12 illustrates the effects of $J_L$ on the time-averaged radial velocity in radial distribution at constant $J_G = 0.0096$ m/s and L/D = 60. At the lowest $J_L = 0.0405$ m/s, the radial velocity profile shows a relatively straight gradient trend with one equilibrium position ($U_{radial} = 0$ m/s) in r/R about 0.56. Equilibrium position in r/R is defined as the radial distribution position where the time-averaged velocity value amounts zero. This may imply that in the equilibrium position the total magnitude of the velocity toward the pipe wall is equal to the total magnitude of the velocity toward opposite direction (toward centre of pipe). For the lowest $J_L$ in Fig. 12, the pattern illustrates that the bubbles in range r/R = 0 - 0.56 tend to move toward the centre of pipe (positive radial velocity), while at r/R = 0.56 – 1 the bubbles are likely to move toward the pipe wall (negative radial velocity).

Fig. 13 shows the bubble size distribution samples for the lowest $J_L$ in Fig. 12. The small bubbles with diameter less than 6 mm are commonly found in r/R = 0.75 – 0.8. On the other hand, for r/R = 0 – 0.05, the bubbles larger than 6 mm-diameter are observed. This indicates that the small bubbles less than 6 mm-diameter tend to move toward the pipe wall. This is consistent with the experimental results of Lucas et al. [27] and Tomiyama [24]. Tomiyama [24] showed that small bubble below 5.6 mm-diameter has a positive lift force coefficient, thus has negative lift force which act on the bubbles toward the pipe wall.

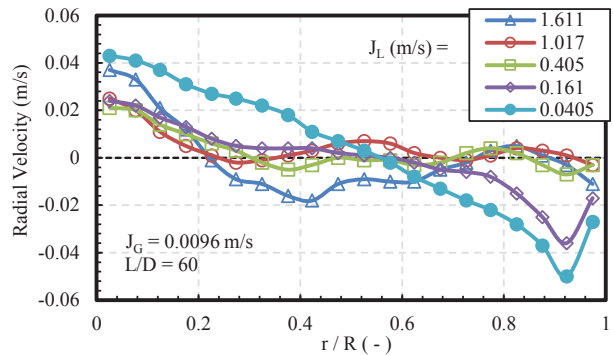

Fig. 12. The effects of $J_L$ on the time-averaged radial velocity in radial distribution.

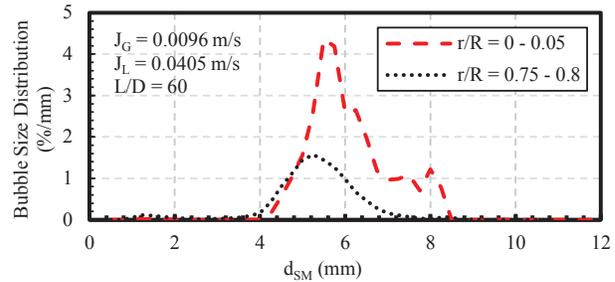

Fig. 13. Bubble size distribution for specific radial distribution range.

As shown in Fig. 12, for $J_L < 1$ m/s, a steep increase profile near the pipe wall (r/R = 0.9 - 1) is noted. This is interpreted when the small bubbles move toward the pipe wall due to the lift force, the wall force gives an opposite force to the bubbles, resulting the decreasing radial velocity magnitude of the bubbles. This must be remembered that the negative value in the time-averaged diagram shows the direction of the velocity. The decreasing velocity of the moving bubbles in radial direction is shown by the shifted velocity value get near to 0 m/s. For $J_L \geq 1$ m/s, the wall force effect is not observed. This is because the wall forces acting on the bubbles are by far weaker than the dispersion force at the given $J_L$. The dispersion force is the result of the turbulent fluctuations of liquid velocity [28]. Hence, the dispersion force will increase if $J_L$ increases.

In Fig. 12, as the $J_L$ increases, the straight gradient trend becomes more flattened with a tendency to be the wavy-like profile. As seen in Fig. 14, the wavy-like radial velocity distribution profile is characterized by the presence of the bends. The number of bends increases relatively if $J_L$ increases, too (see Fig. 12). This indicates an increasing fluctuation of the bubbles radial velocity directions. The dispersion force disperses the bubbles out of the pipe centre. Meanwhile, as the bubble growing along the pipe and tend to move to the centre of the pipe because the lift force, the dispersion force act to give an opposite force. Thus, this interaction will result a fluctuation in the bubbles as the bubbles move upward along the liquid flow.

Fig. 14 displays the location of the bends in the radial velocity profile for $J_L = 1.017$ m/s and $J_G = 0.0096$ m/s at $L/D = 60$. The bends are located about $r/R = 0.28; 0.52; 0.72; 0.84$. The bubble size distributions on the $r/R$ of the bends location are shown in Fig. 15. The bimodal bubble size distribution is appeared on the bends location. This may imply that, on this point, two dominant bubbles classes have opposite orientation direction, resulting a significant resistance to the opposite bubbles direction. Hence, the averaged radial velocity at that point starts to change its magnitude toward equilibrium position (0 m/s). This can be inferred that the bends indicate the presence of the bubbles that change the dominant bubbles direction, resulting a thrust to shift the dominant direction to the opposite direction (for example from direction toward the pipe wall to direction toward the centre of pipe).

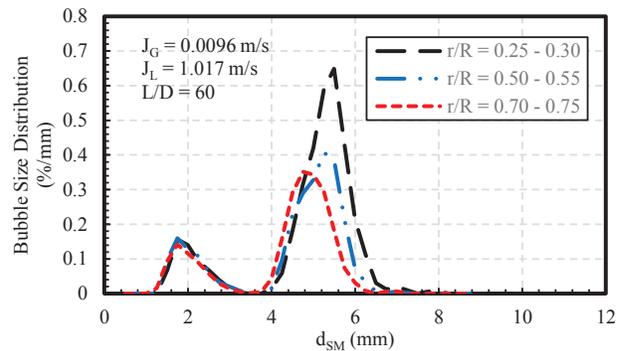

Fig. 15 Bubble size distribution on the bends location.

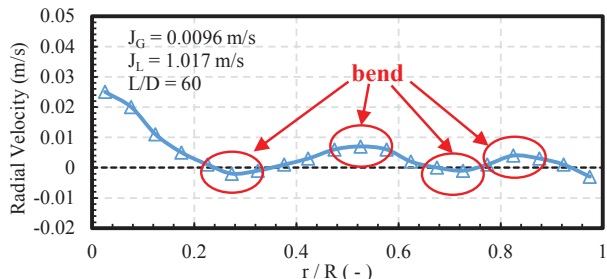

Fig. 14. The wavy-like time-averaged radial velocity distribution profile.

Fig. 16 shows the effects of $J_G$ on the time-averaged radial velocity profile in radial distribution. The wavy-like velocity distribution profiles are observed for all cases. This is because the given flow regimes have $J_L > 1$ m/s, which relatively has high dispersion force resulting in high dispersal movement of the bubbles. For the case of the highest $J_G = 0.0898$ m/s, the relatively high negative radial velocity is observed in the range near the pipe wall. As the $J_G$ increases, the void fraction also increases, the number of the dispersed bubbles increases, too. The increasing number of dispersed bubbles will increase the number of the bubbles that are dispersed toward the wall by the liquid dispersion force, resulting the comparatively high negative averaged value of the radial velocity. Fig. 17 strengthens this interpretation. This is shown from Fig. 17 that the distribution of the bubbles with diameter less than 6 mm are mostly observed near the pipe wall in the highest $J_G$ in Fig. 16, resulting a high number of the dispersed bubble toward the pipe wall.

Fig. 18 and 19 illustrate the effects of $J_L$ and $J_G$ on the time-averaged azimuthal velocity distribution profile, respectively. The time-averaged azimuthal velocity distribution profiles illustrate the interaction effects between lateral forces in angular direction. The wavy-like velocity distribution profiles are observed for all cases in Fig. 18. The increasing value of $J_L$ does not give a remarkable change in the azimuthal velocity distribution profiles. This is contrary to the radial velocity distribution profiles in Fig. 12 that the profiles gradually change from a straight gradient distribution profile to be a wavy-like distribution profile as the $J_L$ increases.

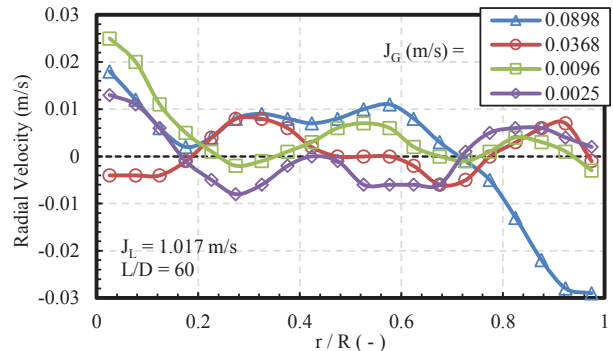

Fig. 16. The effects of $J_G$ on the time-averaged radial velocity in radial distribution.

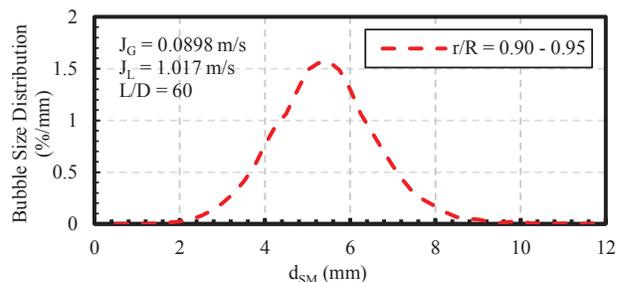

Fig. 17. Bubble size distribution for the highest $J_L$ in Fig. 16.

On the other hand, in Fig. 19, the azimuthal velocity distribution profiles change from a relatively wavy-like distribution profile to be a sharp-bend distribution profile as the $J_G$ increases. The sharp-bend velocity distribution profile is characterized by the appearance of a bend with a sharp-turn side near the pipe wall position ($r/R = 1$), as shown in Fig. 20. The bend position shows the maximum averaged azimuthal velocity that can be reached at given flow regimes. If the $J_G$ increases, the void fraction and the dispersed gas bubbles in the liquid flow will increase as well. The increasing amount of the dispersed bubbles in the liquid flow creates a tendency movement to the clockwise direction.

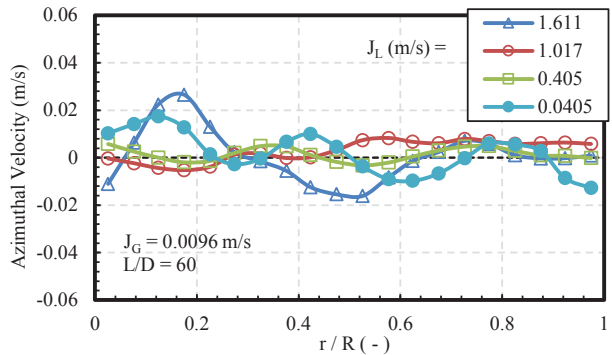

Fig. 18. The effects of $J_L$ on the time-averaged azimuthal velocity in radial distribution.

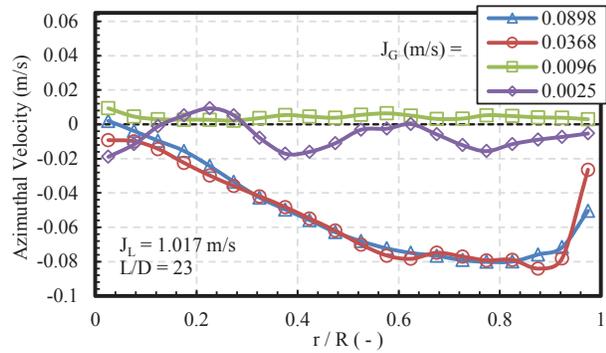

Fig. 19. The effects of $J_G$ on the time-averaged azimuthal velocity in radial distribution.

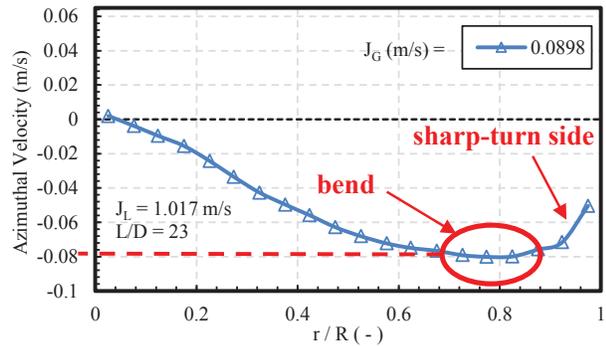

Fig. 20. The sharp-bend time-averaged azimuthal velocity distribution profile.

In Fig. 19, for $J_G > 0.03$ m/s, the gas dispersed bubbles will have a higher averaged azimuthal velocity magnitude as the bubbles are located away from the centre of pipe. Then, the bubbles will reach its maximum averaged azimuthal velocity in the bend position. After that, the averaged azimuthal velocity decreases significantly in magnitude as the bubbles are located near the wall as influence by the pipe wall shear stress. This interpretation is consistent with observations by Moursali et al. [29], which indicated that the bubbles sliding along the pipe wall generated significant velocity perturbations. The effects of the pipe wall shear stress to the bubbles azimuthal velocity are not relatively noted when the $J_G$ is low ($J_G < 0.03$ m/s). This may suggests the reason why there is no notable change in Fig. 18 due to the low $J_G$ in the given flow regime.

## V. CONCLUSIONS

To study the three dimensional bubble velocities at co-current gas-liquid vertical upward bubbly flow, a bubble pair algorithm had been implemented to assign correct paired bubbles from both measurement layers of the ultrafast dual-layer electron beam X-ray tomography ROFEX. The bubble velocity calculations from the bubble pair algorithm's results show satisfactory agreement with the available experimental database and theory. Therefore, the bubble pair algorithm from Patmonoaji et al. [16] is reliable to implement.

From the bubble pair algorithm results, the calculation for the three dimensional bubble velocities can be done. The results point out that lift-, wall- and dispersion-forces play an important role in the development of three dimensional bubble velocity distribution profiles. The interplay between these forces is used to explain the characteristics of the distribution velocity profile. A deep understanding about these lateral forces can be the key to predict the velocity profile for the bubbly flow cases. The present results are also consistent with previous work.

ACKNOWLEDGMENT

The first author would like to acknowledge Institute of Fluid Dynamics, Helmholtz-Zentrum Dresden-Rossendorf (HZDR), Germany for the financial support during 4 months research internship at HZDR.

The first author wishes to acknowledge the Fast Track Scholarship Program from Indonesia Directorate General for Higher Education (DIKTI), Ministry of Education and Culture, Republic of Indonesia.